\input harvmac\skip0=\baselineskip


\lref\fw{D.~Freed, E.~Witten, ``Anomalies in String Theory with
D-Branes,'' [arXiv:hep-th/9907189]}

\lref\DG{M.~Douglas and G.~Moore, ``D-branes, Quivers, and ALE
  Instantons,''
[arXiv:hep-th/9603767]}

\lref\Denf{F.~Denef, ``Quantum Quivers and Hall//Hole Halos,''
JHEP{\bf 10}, 023 (2002)[arXiv:hep-th/0206072]}

\lref\Gep{A.~Recknagel and V.~Schomerus, ``D-branes in Gepner models,''
Nucl. Phys. {\bf B531}, 185-225 (1998) [arXiv:hep-th/9712186]}

\lref\Kth{R.~Minasian and G.~Moore, ``K-theory and Ramond-Ramond
Charge,'' JHEP {\bf 11}, 002 (1997) [arXiv:hep-th/9710230] }

\lref\Wit{E.~Witten, ``D-branes and K-theory,'' JHEP {\bf 9812}, 019
(1998) [arXiv:hep-th/9810188]}

\lref\Har{M.~B.~Green, J.~A.~Harvey and G.~W.~Moore, ``I-brane inflow
and anomalous couplings on D-branes,'' Class Quant. Grav. {\bf 14}, 47
(1997) [arXiv:hep-th/9605033]}

\lref\Che{Y.~K.~Cheung and Z.~Yin, ``Anomalies, branes, and currents,
'' Nucl. Phys. {\bf B517}, 69 (1998) [arXiv:9710206] }

\lref\WiPh{E.~Witten, ``Phases of $N=2$ Theories in Two
Dimensions,'' Nucl. Phys. {\bf B403}, 159-222 (1993) }

\lref\Bru{I.~Brunner, M.~R.~Douglas, A.~Lawrence, C.~Romelsberger,
``D-branes on the Quintic,'' JHEP {\bf 0008}, 015 (2000)[arXiv:hep-th/9906200]}
\lref\aspone{
  P.~S.~Aspinwall,
  ``D-branes on Calabi-Yau manifolds,''
  [arXiv:hep-th/0403166].
}
\lref\asptwo{
  P.~S.~Aspinwall and S.~Katz,
  ``Computation of superpotentials for D-Branes,''
  [arXiv:hep-th/0412209].
}

\lref\aspthree{
  P.~S.~Aspinwall and L.~M.~Fidkowski,
  ``Superpotentials for quiver gauge theories,''
  [arXiv:hep-th/0506041].
}

\lref\aspfour{P.~S.~Aspinwall and A.~Lawrence,``Derived Categories
and Zero-Brane Stability,''
[arXiv:hep-th/0104147].}

 \lref\mrdthree{
  M.~R.~Douglas, S.~Govindarajan, T.~Jayaraman and A.~Tomasiello,
  ``D-branes on Calabi-Yau manifolds and superpotentials,''
  Commun.\ Math.\ Phys.\  {\bf 248}, 85 (2004)
  [arXiv:hep-th/0203173].
}
\lref\mrdtwo{
  M.~R.~Douglas,
  ``Lectures on D-branes on Calabi-Yau manifolds,''
{\it Prepared for ICTP Spring School on Superstrings and Related Matters, Trieste, Italy, 2-10 Apr 2001}
}\lref\mrdone{
  I.~Brunner, M.~R.~Douglas, A.~E.~Lawrence and C.~Romelsberger,
  ``D-branes on the quintic,''
  JHEP {\bf 0008}, 015 (2000)
  [arXiv:hep-th/9906200].
}

\lref\GreeneDenef{ F.~Denef, B.~R.~Greene and M.~Raugas, ``Split
attractor flows and the spectrum of BPS D-branes on the quintic,''
JHEP {\bf 0105}, 012 (2001) [arXiv:hep-th/0101135].
}
 \lref\quinticpaper{
  D.~Gaiotto, M.~Guica, L.~Huang, A.~Simons, A.~Strominger and X.~Yin,
  ``D4-D0 branes on the quintic,''
  arXiv:hep-th/0509168.
}

\lref\Hellermanone{
  S.~Hellerman and J.~McGreevy,
  ``Linear sigma model toolshed for D-brane physics,''
  JHEP {\bf 0110}, 002 (2001)
  [arXiv:hep-th/0104100].
}

\lref\Hellermantwo{
  S.~Hellerman, S.~Kachru, A.~E.~Lawrence and J.~McGreevy,
  ``Linear sigma models for open strings,''
  JHEP {\bf 0207}, 002 (2002)
  [arXiv:hep-th/0109069].
}

\lref\Asen{A.~Sen,``SO(32) spinor of type I and other solitons on
brane-antibrane pair,''
  JHEP {\bf 9809}, 023 (1998)
  [arXiv:hep-th/9808141].
}
\lref\DenefRU{
  F.~Denef,
  JHEP {\bf 0210}, 023 (2002)
  [arXiv:hep-th/0206072].
}

\lref\FDenef{F.~Denef,``D-brane ground states, multicenter black
holes, DT/GW correspondence, and the OSV conjecture, '' "Black
holes, topological strings, and invariants of holomorphic
submanifolds" conference at Harvard, January 2006 }

\lref\DouglasGI{
  M.~R.~Douglas,
  ``D-branes, categories and N = 1 supersymmetry,''
  J.\ Math.\ Phys.\  {\bf 42}, 2818 (2001)
  [arXiv:hep-th/0011017].
}

\lref\SharpeQZ{
  E.~R.~Sharpe,
  ``D-branes, derived categories, and Grothendieck groups,''
  Nucl.\ Phys.\ B {\bf 561}, 433 (1999)
  [arXiv:hep-th/9902116].
}

\lref\GovindarajanEF{
  S.~Govindarajan, T.~Jayaraman and T.~Sarkar,
  Nucl.\ Phys.\ B {\bf 593}, 155 (2001)
  [arXiv:hep-th/0007075].
}

\lref\GovindarajanVI{
  S.~Govindarajan and T.~Jayaraman,
  Nucl.\ Phys.\ B {\bf 600}, 457 (2001)
  [arXiv:hep-th/0010196].
}

\lref\GovindarajanKR{
  S.~Govindarajan and T.~Jayaraman,
  Nucl.\ Phys.\ B {\bf 618}, 50 (2001)
  [arXiv:hep-th/0104126].
}

\Title{\vbox{\baselineskip12pt}}{D4-branes on Complete
Intersection in Toric Variety}

\centerline{Davide Gaiotto and Lisa Huang}
\smallskip
\centerline{Jefferson Physical Laboratory, Harvard University,
Cambridge, MA 02138} \vskip .6in \centerline{\bf Abstract} {We
consider D4-branes on toric Calabi-Yau spaces. The quiver gauge
theory that describes several D4-branes on the Calabi-Yau has a
Higgs branch, that describes configurations of a single large
D4-brane with the same charges. We propose that the world volume
of such a D4-brane is described by a determinantal variety. We
discuss a description of the Higgs branch of the moduli space in
terms of a quiver with twice as many nodes and only bifundamental
fields, arising from a $D6-\overline{D}6$ system.  We recast the
tachyon condensation of the $D6-\overline{D}6$ system in the
language of open string gauge linear sigma model. } \vskip .3in

\smallskip
\Date{December 29, 2006}

\newsec{Introduction}
BPS black holes in IIA compactification on a Calabi-Yau have a
dual description in terms of wrapped D-branes.  The world volume
theory of D-branes wrapping different cycles of the Calabi-Yau is
a quiver gauge theory with each node corresponding to the wrapped
D-branes and bifundamental fields corresponding to massless open
strings stretched between them
\refs{\DG\mrdone\Denf\GreeneDenef\mrdtwo\mrdthree\aspone
\asptwo\aspthree-\quinticpaper}. Ultimately, it would be
interesting to understand the microscopic degeneracies of BPS
black holes from features of the moduli space of the gauge theory.
Toward this end, in \quinticpaper\ we studied an example where we
have N D4-branes and M D0-branes on the quintic.  We proposed that
the quantum mechanics of N D4-branes and M D0-branes on the
quintic is described by a dimensional reduction of a $U(N)\times
U(M)$ quiver gauge theory with the appropriate superpotential. We
showed that the moduli space of the Higgs branch reproduces the
moduli space of hypersurfaces with a certain flux on it.  In this
note, we generalize the classical geometry analysis in
\quinticpaper\ for toric Calabi-Yau spaces. In section 2, we again
find that the Higgs branch of N D4-branes is described by a
determinantal variety. In section 3, we revisit the construction
of the D4-brane moduli space in \quinticpaper\ and recover the
D4-brane moduli space from tachyon condensation of the
$D6-\overline{D6}$ system \Hellermanone \Hellermantwo. In section
4, we comment on the open string gauged linear sigma model
description of D-branes. We recover our determinantal equation for
the D4-branes by solving the F-term constraint in this formalism.

\newsec{Generalizing the quintic story}

The construction developed for the quintic can be generalized
straight forwardly to Calabi-Yau manifolds arising from complete
intersections in toric varieties.

Let us consider a toric variety constructed as a sympletic
quotient ${\cal T}={\bf C}^m // {\bf C}^{*n}$. Let $z_i$ be the
coordinates on ${\bf C}^m$, and the ${\bf C}^*$'s acts by
\eqn\cstar{ z_i \to \lambda^{q_\alpha^i}
z_i,~~~~\alpha=1,\cdots,n. } The quotient space has $n$ Kahler
moduli, corresponding to blown up ${\bf P}^1$'s. In physical terms
they are controlled by Fayet-Illiopoulos parameters in the D term
constraints for the corresponding gauged linear sigma model \WiPh.
The Calabi-Yau $X$ is given as a complete intersection
\eqn\cy{P_1(z)=0,~~\cdots ,~~ P_{m-n-3}(z)=0,} where the $P_{k}$
are polynomials in $z$ that are homogeneous with respect to the
${\bf C}^*$ actions, i.e. they scale as \eqn\sc{P_k(z)\to
\lambda^{Q_{\alpha}^k}P_k(z)} under \cstar. The vanishing of
$c_1(X)$ requires \eqn\cycd{\sum_{i}q_{\alpha}^i =
\sum_{k}Q_{\alpha}^k} for all $\alpha$.

The toric variety ${\cal T}$ is naturally equipped with $n$ line
bundles ${\cal L}_\alpha$, whose chern classes
$x^{\alpha}=c_1({\cal L}_\alpha)$ restrict to the Calabi-Yau $X$
to give the generators of $H^{1,1}(X)$. We will denote by
$J^\alpha$ the corresponding integral harmonic $(1,1)$-forms. A
holomorphic surface $P$ in the Calabi-Yau can be described by one
polynomial equation \eqn\four{A(z)=0} For generic $A$, the only
integral harmonic $(1,1)$ forms on $P$ are the $J^{\alpha}$'s.

As in the last section of \quinticpaper, we want to consider a set
of D4-branes, labelled by an index $I$, wrapped on the cycles
$n_{\alpha}^{I}$ and carrying gauge field flux $F^I = k_{\alpha}^I
J^{\alpha}$. We will give the precise description of the
corresponding quiver gauge theory in the next section, and justify
some of the following claims. It is natural to propose that the
world volume of the D4-brane corresponding to a generic point on
the Higgs branch of the quiver gauge theory is described by the
following equation \eqn\specly{\det [A_{IJ}(z)]=0 } Here
$A_{IJ}(z)$ are polynomials in $z_i$'s with homogeneous scaling
\eqn\homsc{ A_{IJ}(z) \to \lambda^{{{n_{\alpha}^I + n_{\alpha}^J}
\over 2} + k_{\alpha}^I - k_{\alpha}^J} A_{IJ}(z) } under \cstar.
Such surfaces are special in that they contain an extra integral
harmonic $(1,1)$ form dual to the curve $C$ given by
\eqn\curv{C:~~~\sum_{I} v_{I}(z) \tilde A_{IJ}(z)=0} where $\tilde
A_{IJ}(z)$ are minors of the matrix $A(z)$ and $v(z)$ are
polynomials in $z$ of homogeneous degree $\half n^I_\alpha +
k^I_\alpha$. Note that $C$ is not a complete intersection and is
not homologous to linear combinations of $J^\alpha$'s.

As in \quinticpaper, charge conservation requires an appropriate
flux on the D4-brane world volume $P$. It is given by \eqn\flu{F =
C - {1 \over 2}\sum_{I}n_{\alpha}^I J^{\alpha} } To verify the
agreement of D2 and D0-brane charges one needs to compute the
intersection numbers $C \cdot J^{\alpha}$ and $C \cdot C$, along
the same lines as for the quintic. The final expressions are
\eqn\inter{\eqalign{& C \cdot J^\alpha = C^{\alpha \beta \gamma}
\left[{1 \over 2}(\sum n^I_{\beta})(\sum n^I_{\gamma}) + \sum
n^I_{\beta} k^I_{\gamma} \right], \cr & C \cdot C = C^{\alpha
\beta \gamma}\left[{1\over 6} (\sum n^I_{\alpha})(\sum
n^I_{\beta})(\sum n^I_{\gamma}) + {1\over 12 }\sum
n^I_{\alpha}n^I_{\beta}n^I_{\gamma} \right. \cr &
~~~~~~~~~~~~~~~~~~~~~~\left. + \sum n^I_{\alpha}\sum
n^J_{\beta}k^J_{\beta} + \sum n^I_{\alpha}k^I_{\beta}k^I_{\gamma}
\right]. } } where $C^{\alpha\beta\gamma}=\int J^{\alpha}\wedge
J^{\beta} \wedge J^{\gamma}$.  As we will see the line bundle with
curvature $F$ is essentially the bundle of zero eigenvectors of
$A_{IJ}$.

\newsec{D4-brane moduli space from tachyon condensation}

\subsec{The quintic}

In \quinticpaper\ the moduli space of a stack of N D4-branes
wrapped on a hypersurface in the quintic with minimal flux on each
was obtained by solving the D-term constraints of a $U(N)$ gauge
theory with four adjoint chiral matter.  It is well known from
Sen's work \Asen\ that a D4-brane can be realized as a
topologically nontrivial tachyon configuration on the
$D6-{\overline{D6}}$-brane system with the appropriate fluxes on
it.  On a Calabi-Yau manifold, the tachyonic open string going
between a D6-brane and a $\overline{D6}$-brane with different
$U(1)$ gauge fluxes can condense and give rise to a D4-brane,
whose charge is determined by the fluxes. Let us consider the case
of the quintic. A D4-brane wrapped on a degree $d$ surface D, and
with world volume flux $k J$ can be interpreted as a (derived
category of) coherent sheaf ${\cal O}_D$, with locally free
resolution \aspone\aspfour\DouglasGI\SharpeQZ \foot{ $k$ is an
integer or a half-integer depending on whether $d$ is even or odd,
due to Freed-Witten anomaly.\fw } \eqn\locaf{ 0\to {\cal O}(k-{d
\over 2}) \longrightarrow^{\!\!\!\!\!\! \!\! T}~\, {\cal O}(k+ {d
\over 2})\to {\cal O}_D \to 0  } where ${\cal O}(k-{d\over 2})$
and ${\cal O}(k+{d\over 2})$ are the gauge bundle of a
$\overline{D6}$-brane and a D6-brane with $k-{d\over 2}$ and
$k+{d\over 2}$ units of fluxes, respectively.  The zero modes of
the open string ``tachyons" T, are given by the holomorphic
sections of the gauge bundle ${\cal O}(d)$, that is they are
polynomials in $z_i$ of homogeneous degree $d$, up to multiples of
the quintic equation.

The D4-brane world volume is at the zero locus of the tachyon
profile. We would like to propose a supersymmetric quiver gauge
theory describing the bound state of the brane/anti-brane system,
with gauge group $U(1)\times U(1)$ and bifundamental matter $T_a$
coming from the tachyons, where $a=1,\cdots,\dim H^0(X,{\cal
O}(d))$. There is no gauge invariant superpotential for this
theory since there are no loops in the quiver diagram. Solving the
D-term constraints give a moduli space of vacua which is the
projectivization of $H^0(X,{\cal O}(d))$. This matches the moduli
of the degree $d$ surface.  A similar discussion can be found in
\FDenef.

It might appear unexpected to have a supersymmetric quiver theory
that describes the $D6-\overline{D6}$ system. However, from the
four dimensional point of view they are two BPS objects with
generically different central charges. Along a wall in the
K\"ahler moduli space of $X$ they can be mutually supersymmetric,
i.e. their central charges are aligned. Deforming the K\"ahler
moduli corresponds to turning on FI parameter in the gauge theory,
and then the bifundamental open string fields become tachyonic.

Similarly we can describe the quiver theory for a stack of $N$
D4-branes wrapped on a degree $d$ surface. This is a $U(N)\times
U(N)$ gauge theory with bifundamental matter $T_a$ in $({\bf N},
\overline {\bf N})$. Again there is no superpotential and the
moduli space of vacua is given by a symplectic quotient of the
$H^0(X,{\cal O}(d))^{\oplus N^2}$ by $GL(N)\times
GL(N)$.\foot{This is an $N \times N$ matrix with whose elements
are in $H^0(X,{\cal O}(d))$.}  Now the D4-brane is given by the
sheaf of the zero eigenvector of the tachyon matrix. It is
supported on the locus $\det \left(T_{(A_1\cdots A_N)}
z^{A_1}\cdots z^{A_N}\right)=0$. This is precisely the special
surface corresponding to $N$ joined D4-branes as in \quinticpaper.

\subsec{Toric Calabi-Yau Threefolds}

Let us now return to the case of the Calabi-Yau $X$ given by a
complete intersection in toric variety ${\bf C}^m//{\bf C^*}^n$ as
in section 2. A set of $N$ D4-branes wrapped on cycles
$n_\alpha^I$ with fluxes $k^I_\alpha J^\alpha$ can be described in
terms of a complex of vector bundles, as a $D6-\overline{D6}$
system \eqn\dsdssw{0\to \bigoplus_{J=1}^N \bigotimes_\alpha {\cal
L}_\alpha^{k_\alpha^J-{n_\alpha^J\over 2}}
\longrightarrow^{\!\!\!\!\!\! \!\! T}~\, \bigoplus_{I=1}^N
\bigotimes_\alpha {\cal L}_\alpha^{k_\alpha^I+{n_\alpha^I\over 2}}
\to 0 } The tachyon $T$ has components $T_{IJ}$ living in the line
bundle $\bigotimes_\alpha {\cal L}_\alpha^{{n_\alpha^I
+n_\alpha^J\over 2} +k_\alpha^I-k_\alpha^J}$. The quiver theory
has gauge group $U(1)^N\times U(1)^N$ and bifundamental fields
$T_{IJ}$. Again, the D4-brane is given by the zero eigenvectors of
 $T$, which is indeed supported along the surface \specly. The
moduli space of the D4-branes is again determined by the D-term
constraints of this quiver theory.

Starting from the derived category description of D-branes, a
quiver gauge theory can be built along the same lines from the
physical setup of several different $D6$ and $\overline{D6}$
branes. These more complicated setups will generally allow for
superpotentials. Understanding the properties of these quiver
gauge theories may lead to an account of the microscopic
degeneracies of BPS black holes.

\newsec{Gauged linear sigma model}

The world sheet theory of closed strings on the Calabi-Yau space
$X$ can be described as the IR limit of a $(2,2)$ gauged linear
sigma model with gauge group $U(1)^n$ and chiral fields $Z^i$ and
$\Lambda^k$, where $Z^i$ are assigned charges $q_\alpha^i$ under
the $\alpha$th $U(1)$ \WiPh. $\Lambda^k$'s are Lagrangian
multipliers of charge $-Q_\alpha^k$ so that the superpotential
\eqn\suppo{ W = \sum_k \Lambda^k P_k(Z^i) } is gauge invariant.

Following \Hellermanone\Hellermantwo\foot{Also see
\GovindarajanEF\GovindarajanVI\GovindarajanKR.}, the D4-brane
\dsdssw\ can be described by the boundary couplings \eqn\bdysup{
\int_{\partial \Sigma} d\theta\, {\rho}^I T_{IJ}(Z^i) \beta^J }
where $\rho^I$ and $\beta^J$ are bosonic and fermionic fields
living on the boundary with charges $-\half
n_\alpha^I-\kappa_\alpha^I$ and $-\half
n_\alpha^J+\kappa_\alpha^J$, respectively. $T_{IJ}(Z^i)$ are
polynomials in $Z^i$'s with homogeneous charges ${n_\alpha^I +
n_\alpha^J\over2} +\kappa_\alpha^I - \kappa_\alpha^J$.  There is
also a $U(1)$ gauge symmetry acting on the boundary fields
\eqn\uones{ \rho^I \to e^{-i\alpha} \rho^I,~~~~\beta^I\to
e^{i\alpha} \beta^I }. This is the world sheet analogue of the
previous section of building D4-branes from D6-$\overline{D6}$
branes.

The F-term constraints set $\rho^I T_{IJ}=0$. Additionally,
projecting onto charge one states of the boundary theory requires
$\rho$ to be nonzero and therefore \eqn\solsac{ \det T(Z^i)=0 }
This again is the geometry of the D4-brane in the Higgs branch.
Generically $T$ has one zero eigenvector $\beta_0^I(Z)$,
\eqn\rhoss{ T_{IJ} \beta_0^J = 0 } $\beta_0^I$ spans a sub-line
bundle ${\cal L}$ of $\bigoplus_I {\cal L}_I$, i.e. ${\cal L}={\bf
Ker}\, T$. It is clear that ${\cal L}$ is dual to $C-\sum
n_\alpha^IJ^\alpha=F-\half \sum n_\alpha^IJ^\alpha$.

\bigskip

\centerline{\bf Acknowledgements} We are grateful to Xi Yin for
his many inputs and discussions.  We thank Allan Adams for his
discussion of the gauge linear sigma models.

\listrefs

\end